# Knowledge Isn't Power

The Ethics of Social Robots and the Difficulty of Informed Consent

James M. Berzuk
Department of Computer Science
University of Manitoba
Winnipeg, Canada
berzukj@myumanitoba.ca
0000-0003-0242-4993

Lauren Corcoran
Faculty of Law
University of Manitoba
Winnipeg, Canada
corcoral@myumanitoba.ca
0009-0005-0167-4351

Brannen McKenzie-Lefurgey
Faculty of Law
University of Manitoba
Winnipeg, Canada
mckenzib@myumanitoba.ca
0009-0003-2231-1483

Katie Szilagyi
Faculty of Law,
University of Manitoba
Winnipeg, Canada
Katie.Szilagyi@umanitoba.ca
0000-0003-4821-1504

James E. Young
Department of Computer Science
University of Manitoba
Winnipeg, Canada
young@cs.umanitoba.ca
0000-0002-6132-0118

## Abstract

Contemporary robots are increasingly mimicking human social behaviours to facilitate interaction, such as smiling to signal approachability, or hesitating before taking an action to allow people time to react. Such techniques can activate a person's entrenched social instincts, triggering emotional responses as though they are interacting with a fellow human, and can prompt them to treat a robot as if it truly possesses the underlying life-like processes it outwardly presents, raising significant ethical questions. We engage these issues through the lens of informed consent: drawing upon prevailing legal principles and ethics, we examine how social robots can influence user behaviour in novel ways, and whether under those circumstances users can be appropriately informed to consent to these heightened interactions. We explore the complex circumstances of human-robot interaction and highlight how it differs from more familiar interaction contexts, and we apply legal principles relating to informed consent to social robots in order to reconceptualize the current ethical debates surrounding the field. From this investigation, we synthesize design goals for robot developers to achieve more ethical and informed human-robot interaction.

## Acknowledgements

This projected was funded by the Natural Sciences and Engineering Research Council of Canada (NSERC) through their Discovery Grants program, and the University of Manitoba through the Faculty of Science Interdisciplinary Research Grant. It was further supported by the NSERC Canada Graduate Scholarships – Master's scholarship, the University of Manitoba Graduate Fellowship, and the University of Manitoba Faculty of Science Undergraduate Student Research Award.

# 1   Introduction

Social robots are designed to use human- or animal-like language, behaviours, and emotional modelling to facilitate smooth and effective communication with people, with an expectation that people will naturally understand these socially rooted communication techniques [69]. Customer service robots may present a pleasant face and a smile to communicate that they are approachable [70], or use hesitation gestures [37] when removing plates from a restaurant table to give people time to react. Designing social robots in this fashion explicitly constructs and presents *artificial* life-like characteristics. By virtue of these design choices, social robots appear to possess emotional and social interaction skills, even though they do not possess the internal capability or state being presented (e.g., genuine friendliness, uncertainty). The result is a thin façade (e.g., an on-face smile) to represent a complex construct, serving as a hollow representation of the authentic social interaction being mimicked. In marked contrast, from a user's perspective, social robots trigger their existing social skills and activate biological instincts; even in infancy, humans recognize movement and face-like shapes as signs of a social agent [58]. This activation can precipitate *bona fide* human emotional responses and social tendencies toward emotional reciprocity and conformity with social norms [38], as if they were interacting with a living entity that was experiencing similar activations. In short, the nature and success of social robotics relies on their ability to prompt humans to respond to them as if they possess the complex life-like processes being represented.

This approach means that social robots can be designed to elicit specific responses in people by presenting a hollow façade of biological social systems, such as maintaining a user's eye gaze to prompt the user to feel as though the robot is paying attention, when it is simply tracking a face. Social robots can quite easily create the impression, in the mind of the user, of a complex, socially grounded, and reciprocal social interaction that does not actually exist. This example highlights an underlying problem with human engagement with social robots: social robots are fundamentally deceptive, making it difficult (or perhaps unreasonable) to expect people to understand, internalize, and accept the nature and risks of the interaction [30]. This raises several ethical questions. First, are people able to be sufficiently aware of social robots' abilities to influence their emotions and behaviour, in comparison to their interactions with other people, to the point where they can make rational decisions when interacting with robots? Second, even when people are properly informed, can we expect them to use this knowledge to react rationally in the face of robotic social interaction that triggers entrenched and biological human responses? In this paper, we engage these issues from the broad lens of informed consent, drawing upon legal principles, medical science, and ethics. We analyze the necessity for people to be informed and aware of the facts and implications of interacting with social robots when weighted against the risks. This analysis results in key considerations that designers can employ to achieve more ethical human-robot interactions.

As people are already familiar with a range of persuasive media and technology (e.g., digital agents, social media influence, advertising, etc.), some may expect that interaction with social robots will slot neatly into well-established frameworks and practices for such encounters; interaction with social robots may simply trigger people's standard existing practices and behaviors. However, there is mounting evidence that social robots can elicit stronger reactions than virtual or tele-present interactive agents (e.g., see [17]); perhaps being physically embodied and collocated ties into humans' evolved tendencies to quickly recognize other living beings [59]. There are further arguments that people may perceive robots as a new ontological category, as something both in-between and distinct from human, animal, or object, and ascribe robots with mental states, intent, sociability, and moral regard [26]. Due to these impacts, we offer a reexamination of people's interactions with social robots from an ethical perspective, focusing on their breadth, opacity, and intensity, especially as compared with other agents, media, or technology with which people commonly interact.

One sensational social robotics research thread demonstrates how robots can persuade, pressure, or manipulate people, often resulting in some pragmatic detriment or a negative emotional state [3]. While such works are often scripted and staged to demonstrate an exaggerated effect, using robots and automated behaviors not feasible or plausible in the real world (e.g., see [16]), there are increasing examples demonstrating the power of social robots with simple behaviors in realistic scenarios (e.g., see [5]). For example, a robot may simply change between static culturally-matched scripts based on knowledge of users' cultural backgrounds to manipulate people into giving a robot more time to help it achieve its goals than they would without the adaptation [50]. These manipulations (and the follow-on potential effects) may be well beyond the public's expectations of a social robot's capabilities. It is naïve to expect people to both anticipate and be readily equipped to resist tailored physically-embodied social interactions executed with algorithmic, robotic precision [68]. Relatedly, deception is often cited as being in direct conflict with informed consent, with some demanding that the use of deception requires very careful consideration in any situation where people must be kept fully informed [45]. In this paper, the lens of informed consent offers a mechanism for exploring the complexities of interacting with social robots.

We begin our analysis in Section 2 by exploring the layers which form the complex and unique circumstances of social robot interaction. These layers allow us to focus our conversation on the unique intersection of physicality, sociality, and technology occupied by social robots. Next, in Section 3, we review key legal principles and lenses which undergird the concept of informed consent. To develop a rounded understanding, in Section 4, we recontextualize the current ethical landscape and debates in social robotics from the perspective of informed consent. Following, in Section 5, we synthesize this information to a pair of key considerations for achieving informed consent in human-robot interaction (HRI). Throughout, we leverage a case study to examine a real robot system in the field with respect to these considerations.

Ultimately, we argue that HRI requires a comprehensive understanding of the potential dangers of social robots in relation to reasonable expectations of human users' ability to understand and make conscious decisions about their interactions with social robots. Our work endeavours to provide a foundation to explore these issues.

## 2 Social Robots as a Unique Interaction Context

Interaction with social robots is a unique social phenomenon, different from interacting with other technologies or living entities. This differentiation stems from the intersection of robots' social and physical embodiment with their superhuman capabilities, positioning social robots as a novel kind of interaction entity. Robots act like they are alive, while remaining a technological artifact, complicating social expectations for living things and technological artifacts alike. Better understanding this unique positioning from a human perspective is critical for exploring the issue of informed consent for interacting with this new class of social technologies.

### 2.1 The Impact of Designed Sociality

Social robots are explicitly designed to engage with humans' emotions and social instincts. For example, they can be designed with outwardly human or animal-like features (e.g., humanoid shape, realistic voice, lifelike gaze following, etc. [6, 43]) and exhibit displays of emotion in order to facilitate interaction or achieve a related goal. This is much less common with more conventional technologies that people are familiar with, which are typically designed for more mechanical or direct informational interaction, raising questions about people's knowledge of the techniques that social robots use.

When a person encounters a social robot, the lifelike features encourage anthropomorphism or zoomorphism (animorphism), where the person attributes the robot as having lifelike (e.g., human- or animal-like) characteristics to help them understand it and determine how to interact with it [15, 73]. This process can vary, from serving as a social expedient leveraging existing knowledge to facilitate interaction, to regarding the robot as a social peer and using it to fulfill social needs [15]. As this process may be grounded in an evolved human tendency [72], it may be quite difficult for an individual to overcome.

Robots presenting lifelike abilities and features is a form of deception that may lead users to make incorrect assumptions about a robot's abilities [54]. Despite the lack of genuine substance behind these social interactions, they can have real impacts on users. For example, one social robot pressured individuals into assisting it for nearly 30% longer by utilizing a script that leveraged the user's cultural background in a socially intelligent manner [50], while another social robot persuaded people to disclose intimate information by first divulging its own 'secrets' [38]. A robot named "Boxie", developed by MIT, was able to obtain seemingly unlimited access to individuals' personal information, because it was "adorable and unthreatening" [20]. Robots have also been shown to be able to guide people toward poor decisions, such as pouring orange juice over a plant [49] or taking a wrong turn during an evacuation scenario [47]. These striking reactions highlight the potential dangers of inherently deceptive animorphistic robot design.

Relatedly, interaction with social robots over time can lead to the development of parasocial relationships [39]: unilateral emotional connections that people can form with artificial entities, fictional characters, and media personalities [7]. These interactions create an illusion of reciprocity between the person and robot, in which the latter's behaviour encourages the fiction of a mutual interpersonal connection. The extent to which these parasocial relationships can develop was demonstrated in 2023, when, as a result of a lawsuit against its parent company, the AI companion chatbot "Replika" had its ability to engage its users in erotic conversation terminated [13]. The outcry from Replika's users was so severe that the company felt that they had to post information about accessing suicide prevention help [13], and ultimately restored the chatbot's erotic capabilities to users whose "relationship" with their Replika began prior to the lawsuit [13].

These connections, though illusory, can have real tangible effects on people. The effects of parasociality may be positive, such as improving a person's perceived sense of well-being [39]. These effects can also be ethically hazardous, as they have been shown to influence a person's spending habits, promoting intent to make a purchase [24] and stimulating impulse-buying behaviour [71]. As parasociality emerges from relationships where a person is engaging in a social 'relationship' without reciprocity, social robots are therefore prime candidates to abuse the power of parasociality to leverage and manipulate users' behaviour. Even when properly informed about the nature of the robot, people may be unable to resist forming such a relationship, just as they form relationships with film or TV characters while rationally aware that those characters do not actually exist. This is especially concerning for children, who have been shown to consider robots as to be social beings and even friends, telling robots secrets that they would not share with an adult [20]. While such risks also exist when dealing with other people, this influx of sociality into traditionally non-social machines allows for social manipulation to emerge from places people might not expect or have experience dealing with.

## 2.2 The Impact of Physicality

A defining trait of social robots is their physicality [21], as social robots dynamically occupy space in the user's environment.

Often capable of moving autonomously, social robots garner a heightened sense of presence and attention, explained in part by human instincts toward detecting motion to identify others and assess threats [59]; even from infancy, humans have been observed to view lifelike patterns of movement as indicative of a social agent. For example, infants respond to robots as social agents, following their gaze as they would a person [35]. These anthropomorphizing reactions—to ascribe human motivations to non-human objects—also happen when observing moving objects with no otherwise-lifelike characteristics. For example, people assign agency and emotional state to the motion of a collection of abstract shapes [22], and even to a stick that simply moves in regular or irregular patterns [19]. While the exact mechanisms behind this phenomenon are not well understood [10], this evidence suggests that the effect is a natural occurrence, grounded in biology.

While virtual on-screen agents garner similarly strong reactions (e.g., as in popular media), there is a large body of research that highlights stark differences between how people respond to robots in comparison with more traditional media [21]. For example, studies have demonstrated that participants assign greater social presence to physically present robots [25] and provide them with more "personal space" [2] versus on-screen virtual agents on a monitor, and rate robots more amicably [25] in general. Further, people can be more compliant with the requests of physical robots in comparison to virtual agents [2]. Social robots have the potential to elicit stronger emotional responses than virtual agents [52]; participants express more sympathy for a robot than a virtual agent that was placed in a distressing situation [52]. People even have heightened emotional responses to a co-present robot versus a live video feed of a remote robot [33], an overall effect which may be linked to the greater perceived physical size of collocated robots in comparison to those displayed on a smaller screen, similar to how people respond to taller individuals versus shorter ones [33].

This heightened potential for ascribing robots with agency and emotional reaction precludes us from regarding social robots as posing equivalent ethical concerns to those of virtual social agents. That these human reactions may be biologically rooted raises questions about the difficulty in resisting such reactions, even when a person is properly informed. This complicates questions of informed consent, as users may find themselves unable to make a meaningful decision no matter how informed they are.

### 2.3 Social Robots as Superhuman

Although social robots' design may present them as lifelike entities, their computation and networking power gives them a wealth of superhuman capabilities not outwardly reflected in their designs. This includes the ability to collect incredible amounts of data from sensors or networked sources and process it at incredible speeds, using this to adjust social interactions in real time. While a person might expect a salesperson to leverage surface observations about them to adjust their sales pitch in real time, a sales robot could leverage high-speed cameras and advanced algorithms to monitor facial and body motions, inferring the human's state, while simultaneously analyzing the person's entire public social networking record, to tailor an efficient and personalized pitch. All of this is invisible to the user as there is no outward indication of the robots' internal machinations.

Robots are not bound by the limitations that may be suggested by their designs; for example, while a robot may close its eyes to signal that it is not watching, it can still observe from cameras not in the eyes [27]. Concurrently, robots can be purposefully designed with constraints and operational inflexibilities to limit and steer user actions, as a kind of "dark pattern" [14]; for example, a robot may pass something to a person but not have the physical capacity to take it back, forcing the user to keep it. Another example of these design constraints is the "always-on, always-listening" device [4], which can place limits on users' privacy with surreptitious monitoring practices and unclear explanations of how to limit their ability to track the consumers [4]. These limitations can coerce a user into making choices they would not otherwise make when interacting with a person.

In human-human interactions, there is an expectation that both parties will be similarly vulnerable, engaged in principles of reciprocity that contribute to a feeling of mutual trust [38]. Social robots seek to influence humans but are not limited by corresponding human social features that would cause them to be influenced in return. This violates basic social expectations that are brought to interactions with other people. By allowing an illusion of reciprocity to persist, in which a person may believe that their actions are affecting the robot similarly to how they would a human, social robots betray that trust and mislead users about the true nature of the interaction, compromising the users' ability to make informed decisions. Social robots are not necessarily constrained by the rules and assumptions of human-human relationships [18].

Social robots possess superhuman abilities and can have manipulative designs that are both unclear to people and contradict expectations people may have given the lifelike designs. Accordingly, a social robot presents superficial humanlike characteristics that lead people to map their expectations of humans onto the robot, even though robots are in no way bound to meet these expectations. This gap between the expectations a robot creates and its actual capabilities, which people may not be aware of, raises important ethical concerns about social robot designs that are not found in human-human interaction.

### 2.4 Social Robots Demand Unique Consideration

Taken together, the active design choices directed towards robots' sociality leverage human biological instincts to demand attention and generate emotional response. These phenomena are particularly pronounced with physical robots as compared against virtual agents. Because of their physical presence and mobility, social robots can manipulate users' internal emotional states, establishing long-term feelings of connection and obligation.

At the same time, robots are different from actual living social agents in their ability to draw on computational power and massive quantities of data, and their inability to meet basic social expectations of reciprocal influence and mutual vulnerability. For these reasons, it is important to consider whether people are truly in control when they interact with social robots, and whether requiring informed consent can alleviate that concern.

However, it is important to remember that these problems are not necessarily intrinsically inherent to the robots themselves. A robot is simply a machine; while its design dictates the possibilities of its potential interactions, the associated risks are largely determined by this design in concert with the context in which it is used. We introduce a case study to illuminate our issues using a specific robot design, interaction design, and use case. We target the use case of a Socially Assistive Robot (SAR) for a child with Autism Spectrum Disorder (ASD), specifically implemented using a Softbank NAO robot, such as work by Shamsuddin et al. [53] that investigated the efficacy of such a robot in this case.

Shamsuddin et al. [53] measured the frequency of a child's ASD-associated behaviours when interacting with a robot, relative to their frequency in the classroom environment. In the roughly 15-minute module of social interaction with the robot, the young student showed that he was able to "… dampen his autistic traits compared to his normal behavior in classroom." [53] Further, a greater amount of eye contact was observed between the child and the robot, relative to his teacher in the classroom setting. Ultimately, the researchers found that the study supported their hypothesis that "… the humanoid robot NAO serves as a significant platform to support and initiate interaction in children with ASD." [53]

This example serves as a useful case study, providing several examples of exactly the risks that we target in this paper. For example, the researchers noted the risk of "'permanent attachment' and the preference of robotic contact compared to actual humans," [53] mirroring the propensity for parasocial relationships; this is exacerbated by the fact that extended interaction may be required for interaction benefits to persist [53]. Further, this raises privacy concerns as children may divulge secrets to robots that they would not share with adults [20]. While such interactions may be relatively harmless with inanimate objects such as toys, social robots may be recording interactions (possibly unknown to the child), children may not properly understand social robots' capabilities or limitations, and thus may not truly grasp the fact that the robot is revealing their secrets to other people. Would robot superhuman capabilities be apparent to a child with ASD? We argue that it is unlikely. In sum, this use case highlights many of the aforementioned issues present in social robots in the context of social interactions with children.

## 3 Legal Perspectives on Consent and Human Autonomy

Social robots have inherently deceptive qualities, with certain design characteristics that provide exceptional powers of influence over the people with whom they interact. This degree of influence means it can be difficult for people to fully comprehend the scope of the interaction or react to it in an informed fashion. From a legal perspective, this might be described as a problem of *informed consent*: one where people may have difficulty making informed, intelligent decisions about their interactions with robots, stemming from a lack of understanding or improperly grounded expectations. This raises several questions, including what level of understanding about social robots do people need to be considered sufficiently informed to make legitimate interaction decisions; how one could determine if such a level had been achieved; and who is responsible for ensuring this level of knowledge is provided. To help us engage these questions and provide tools for analysis, we look broadly to the legal realm for examples of how to grapple with analogous issues. The law offers a body of interpretative lenses and information standards: a framework for interpreting the legitimacy of interactions in society.

Ideas surrounding information and knowledge requirements in relation to a person's autonomy (i.e., one's ability to truly make independent and self-determined decisions and actions) are recurring topics in law, providing legal parallels that we can apply to analyze interactions with social robots. For example, negligence law has developed principles to gauge parties' knowledge and consent to determine the legality of interpersonal conduct and its corresponding impact on other people [75]. These principles are the root of informed consent and provide a useful framing and concrete tools that can be applied as an interpretive lens to interactions with social robots.

In this section, we unpack key legal principles regarding the levels of knowledge, transparency, and consent that are required to ground humans' interactions. First, we look at humans' autonomy to freely form relationships and engage with one another. For example, purchase transactions, employment arrangements, and even advertising claims are all reciprocity-based formal obligations that we create (contract law). Further, people have general expectations of what conduct is acceptable from others beyond formal consent relationships. For example, we expect that people will not harm us as we pass them on the street, that doctors will take appropriate care when treating us, and that the food we purchase will have been safely made [78] (tort law). These expectations do not require express agreement; they are foundational assumptions of our interactions within broader society. Through our exploration of these principles, we articulate how these laws of obligations shape our understanding of social robots and notions of consent. Reflecting the authors' knowledge base, this paper is grounded in a Canadian legal perspective. Much of the material discussed, however, draws from the American and British legal systems, and provides a broadly pertinent overview of concepts, although jurisdictions will differ.

### 3.1 Formal Obligations: Contract Law

A contract is a legally binding agreement which requires (1) an intention to be legally bound, and (2) flowing from a *meeting of the minds*, meaning that all parties understand the essence of what they are agreeing to [1]. If any essential aspect of a contract is missing at the time of agreement, even if parties believe they are in agreement, there can be no meeting of the minds and therefore no contract [79]. Therefore, it is genuinely shared understanding that forms the legal basis for a valid agreement. Typically, contract law functions on a "buyer beware" basis, placing significant onus on the individual to inform themselves; however, social robots may step outside these traditional conventions in ways that buyers are both unaware of and unprepared to handle.

If law requires a *meeting of the minds*, how does it understand human-machine contracts? Contract law recognizes that when one party is using a machine as an intermediary to contract (such as an automated parking attendant or a vending machine), there are restrictions on permitting the machine to modify terms of the contract, as compared to in human-human contract formation [29, 80]. This distinction emerges from an awareness that robots may not be able to be reasoned or bargained with (e.g., they are often ill-equipped to parse novel input). With social robots, not only might users not be properly understood, but they might make allowances for robots' behaviour, with the view that they are 'only' machines. In one study, participants were pressed (by a robot) to continue a menial task indefinitely. While they could theoretically stop at any point, nearly half continued to comply with the robot's demands, in spite of both repeated protests to stop and many participants believing the robot was broken or otherwise mistaken [16]. Given the ambiguous place social robots occupy between mere machines and social actors, it may be difficult to position them within the existing legal landscape, as they may leverage both the sociality of social agents and the allowances users may grant to a robot's perceived machine inflexibly to press the user to make decisions they otherwise may not have made.

Another elucidative contract law concept is mistaken identity. In contracts formed without in-person interaction (i.e., through writing), a contract can be void if one party has deceived another as to their identity [81]. When dealing in-person, however, there is a strong legal presumption that people are well-positioned to understand who they are engaging with [82]. In general, contract law privileges this collocated, embodied communication, and imposes greater liability on the deceived party for in-person interactions, where they are in a strong position to judge the circumstances. The concept of mistaken identity demonstrates that while law values knowledge of identity, it strongly presumes this is not an issue in face-to-face interactions. Social robots frustrate this presumption.

Designed sociality and the impacts of social robots' physicality support strong anthropomorphism, which can trouble a person's ability to understand the nature of who (or what) they are speaking to. While the law values knowledge of identity, it places emphasis on a person's responsibility to know who they are contracting with, and what they are agreeing to, particularly during in-person interactions. With social robots, the in-person nature of the interaction cannot be sufficient to assume human users will be informed (nor necessarily able to be informed) of the nature of the robot's algorithms and abilities, particularly when being potentially "blinded" by the animorphism that suggests they are interacting with a living entity. Contract law reveals how social robots disrupt the established ontological categories that the law is prepared to engage with (simply tool or human) and, as we will discuss, calls for fresh consideration from designers.

### 3.2 Implicit Obligations: Tort Law

Tort law concerns the general social obligations people owe to one another, consistent with societal norms. This includes obligations of general caution and to take appropriate care, avoiding danger or harm to others [83]. These duties of appropriate care are seen in the classic example of *Donoghue v Stevenson*, where a woman became distressed after drinking a ginger beer that contained a decomposed snail [78]. In this case, the courts decided the manufacturer owed all consumers a reasonable standard of product safety. This standard enforces certain behaviours as minimum requirements, even where there are no contracts explicitly governing such behaviours. This is the basic principle of tort law: limits are prescribed on autonomy to protect those who have not consented to the consequences of someone's conduct [32]. For social robots, this balance is critical: not only in terms of what the robots are capable of, but also the impact their actions might have on others.

Beyond these basic minimums and prohibitions on conduct, tort law contains affirmative duties. These duties arise out of relationships of control or power imbalance, where one party may have special knowledge or influence over the other (examples of this relationship include parent-child, lawyer-client, doctor-patient, among others). This principle also applies to businesses where they have created a risk to individuals [77]. For example, businesses serving alcohol have a profit motive to sell as much as possible, but high volumes of alcohol consumption puts patrons at risk and lowers their inhibitions; therefore, patrons are owed a legal duty by the business to take steps to ensure their safety [75]. While some may argue that the choice to over-consume alcohol should be an individual's choice, the law recognizes the influence a profit-motivated business may have to encourage and facilitate risky activities; it cannot be left solely to consumers to protect themselves in these scenarios [75]. Through this lens, profit-motivated businesses designing social robots must consider and develop a standard of accountability congruent with the risk they are creating (such as using sociality to lower users' guards and promote over-trust in social robots). This accountability results from first acknowledging the unique positioning of social robots in legal spaces.

Social robots' superhuman capabilities create a strong power imbalance that can be analogized to affirmative duties on behalf of the robots; the question remains whether these duties are best attributed to the robot creators and designers, or to the robot programming itself. This observation is complicated by the fact that we cannot generally expect people to be fully aware of social robots' capabilities, neither cognizant in a specific interaction (thinking about the robot's capabilities), or necessarily educated enough to understand the superhuman capabilities that the robot possesses, further increasing the importance of the affirmative duties.

Humans are not inherently mistaken by the identity of a social robot (i.e., they do not simply regard them as living things [26]), nor are they completely aware (on all conscious and unconscious levels) that they are merely dealing with a machine [65]. As such, people may expect that robots adhere to human social obligations when they do not, and how these expectations operate may be both unpredictable and controversial.

Additionally, designed sociality may frustrate how one determines reasonable expectations in the tort law sphere: social robots' presentation of faces, emotions, personalities, etc., may lead to people believing that a robot is paying attention, is informed of the specific details of a given interaction (e.g., they saw a person take something), or cares, when it really does none of these. Therefore, a person may hold a reasonable expectation based on the robot's presentation, which is not substantiated by the robot's ability, leading to a misunderstanding. Often, robots' abilities are more limited than their social design choices suggest [48, 51].

### 3.3 Consent and Informed Consent

Law's focus on individual autonomy uses consent as the key mechanism for turning unacceptable risk and vulnerability into a personal choice. Consent is the vehicle which "transforms the brawl into the boxing match" [32]. This results in a balancing act between protecting vulnerable parties who may lack understanding or agency and providing people with personal choice. Some of the legal discourse places an onus on people to understand and inform themselves of the particulars of legal situations, while heightened duties persist to protect them from especially dangerous situations.

However, while it is often assumed that merely possessing knowledge is sufficient to produce informed consent and create congruency between users' desires and behaviours [64], understanding plays a key role. The term "informed consent" refers to a particular, specialized standard used in contexts such as research and medicine [45]. Informed consent stems from principles of autonomy in decision-making, insisting that research and medical interventions should not be done to people without their permission—aiming to further root consent in a meaningful understanding of what an individual is consenting to [45, 74, 76]. This duality—the combination of consent *and* information—relates to circumstances where people are entitled to be sufficiently educated or knowledgeable to understand risks involved. Informed consent often requires higher standards of protection than more general legal contexts. The ability to internalize information, then, is a necessary precondition to meeting the bar of informed consent.

Informed consent emerges predominantly in research and medical contexts precisely because of their specialized nature: it is presumed to be beyond the scope of the average person to understand without explanation [84]. Further, the nature of research and medicine means that, even where people are provided information, it may be highly technical and difficult to understand, limiting one's ability to obtain fully informed consent [45]. For example, a patient who consented to a "tissue removal" may not understand that they are having an entire organ removed, because the true nature of the procedure was obfuscated by the language used. This limitation is the opacity of consent [41], and medical scholarship raises concerns about a patient's level of understanding [36]: not only about the information itself, but also about the nature of the informed consent process. Relatedly, some people are misled to believe that consent is a mere formality [64], rather than an integral part of their legal participation. The ability to internalize information becomes a necessary precondition to meet the informed consent threshold.

The informed consent standard is usually imposed in technically complex and/or risky circumstances. Heightened consent requirements ask experts to exercise their knowledge and expertise in good faith, sharing accessible information with those who cannot otherwise have full awareness of the scope of a situation.

### 3.4 Complexities Arising from Social Robots

Legal standards are often assumed to be generalizable, able to transcend contexts for application to new problems. In the case of applying longstanding legal principles to the research domain of social robots, several complexities arise. Our research highlighted the importance of providing clear information to people on the nature of the social robot, who and what they are interacting with, and the details of a transaction or the interaction. It may be assumed that mere presence of knowledge is sufficient to produce informed consent, and there is onus on users to understand what they are interacting, but these assumptions have shortcomings. Even when appropriate information is provided, the nature of robotics technologies is not always easily understood, meaning that relying on people to mobilize this information and decide for themselves is fraught with misunderstanding. Further, while some legal discussions rely on human autonomy, we cannot assume that people can exercise full autonomy when interacting with social robots.

This is especially true given the propensity for social robots to create vulnerabilities in those with which they interact [13]. The generalizability of legal standards is set aside when individuals are particularly vulnerable: contracts between adults and children, or neurotypical adults and people with disabilities can be unraveled; claims of undue influence can set aside late-in-life changes to a will; and children under the age of 13 are subject to special protections for data collection [13]. Typically, the law treats vulnerability as a status, with more of a binary, yes/no application [13]. However, DiPaola and Calo [13] argue that this vulnerability is dynamic, rather than a fixed status, and can be influenced deliberately by design. As such, issues arising from the potential for exploitation of individuals made vulnerable by social robots (perhaps most of us) is an area that the law must recognize as context-dependent, rather than a generalizable standard.

Technological complexity creates a lack of congruency where information is available but can be quite difficult for a person to parse. Similar situations have been noted in data privacy contexts, where user self-management has often failed to materialize outcomes that match users' stated preferences [60]. People's stated desires about how they want their data managed often do not align with their personal decisions about data privacy [60], suggesting a potential misunderstanding of the choices made. This incongruency means that mere information and choice may not be sufficient to help a person obtain meaningful understanding required to enter into agreements, thereby raising questions about users' autonomy in decisions they make with such systems.

A key source of this incongruency is that the design and interaction circumstances of social robots mean they occupy an uncanny social and legal space between human and machine. We do not argue that people are inherently mistaken by the identity of a social robot (i.e., people do not simply regard them as living things [26]), but rather that people are not completely aware (on all conscious and unconscious levels) that they are merely dealing with a machine [65]. This has resulted in misunderstanding abilities [27], user over-trust [67], and heightened ascription of moral judgement, intention, and agency to social robots [65], creating misunderstanding in human-robot interaction. Taken together, the very human tendency to anthropomorphize robots can undercut the validity of legal decision-making *vis-à-vis* robotic stimuli.

For example, consider a HRI study in which participants were asked to perform a series of menial tasks [16]. Participants were informed that they could quit at any time. However, whenever the humans attempted to quit, the robots prodded them to continue—and almost half of participants obeyed! Notably, some participants reported their belief that robots could not respond to novel stimuli as a reason to obey [16], demonstrating clearly the incongruency between lifelike robot (capable of exerting social pressure) and mere machine (rigid, cannot reason).

Reliance on information availability and users' choice and decision-making is central to understanding HRI. The gap between people's knowledge and behaviour highlights the crux of the issue: knowledge is not inherently power. While it is often assumed that the mere presence of knowledge is sufficient to produce informed consent and create congruency between users' desires and behaviours [64], it is not sufficient to place a 'buyer beware' label on social robots and unduly burden users to extricate the tangled effects of social robots on them [60], or to expect people to fully understand what and who they are interacting with, and how. Instead, we posit that the concept of informed consent acts as a lens by which we can bridge these incongruencies and both highlight and outline the need to strengthen principles of human autonomy and agency in interactions with social robots. Because social robots disrupt existing ontological categories [26], it is important to consider what factors may shape people's understanding of them, thereby altering their ability to meet informed consent standards.

### 3.5 Case Study Reflection: NAO Robot for ASD

These legal issues can be exemplified by their application to our case study by Shamsuddin et al. [53] In the case of a parent or guardian accepting the terms of a contract for a social robot to aid their child with ASD, the accepting party may feel like they are obligated to accept the terms, because the service or product for which they are accepting the contract is a necessity. This might be described as a contract of adhesion. In this context, parents are provided with a binary choice: either accept the legal terms of a robot, or not use it at all. Contracts of adhesion are, in essence, "take it or leave it" contracts, wherein the user of a service must accept the terms of a contract as they are, without opportunity for negotiation [31]. They are increasingly common in the modern era, with terms of such contracts becoming increasingly protective of the drafting party [31]. A "take it or leave it" contract may lead to acceptance of terms that release the company from wrongdoing resulting from the relationship between their child and the robot. While some such agreements have been voided for unconscionability [85], courts tend not to interfere with bad contracts [31] due to the primacy of contractual freedom. This concept may well lead to risk-laden outcomes for NAO's users.

Another concern is the risk of power imbalance between the parties, which is especially problematic in the context of a social robot designed for interaction with children with ASD. As discussed in section 3.2, social robots' superhuman capabilities create the possibility of extreme power imbalances, made even more likely by the specific relationship between a NAO robot and a vulnerable child with ASD. Given the power dynamics of this relationship, and the possibility of harms perpetrated by the robot on the child, the issues of liability become even more pronounced. We query whether tort law would appropriately assign liability in the case of a civil wrong perpetrated by a social robot that, through its designed sociality, has created a power imbalance with its user, a child with ASD.

Further, as discussed in section 3.3, the issue of informed consent becomes more pressing in instances where the context makes a user unlikely to fully understand the agreement they are making, often in highly technical circumstances [46]. The use of the NAO robot would very likely qualify as just such a context, as these kinds of social robots can possess "superhuman" characteristics that are not only unknown to their users, but also difficult for their users to understand, even with a fulsome explanation. As such, the premise of informed consent may be merely a convenient legal fiction in the use of the NAO robot.

A similar inconsistency may be found in the inappropriateness of generalized laws in the case of vulnerable individuals. Children with ASD will almost certainly be more vulnerable to influence from the NAO robot, and laws that are general in nature may be ill-suited to protecting them. In such cases, there must be concessions made in order to avoid these children being put in situations where statutory and common law protections are inadequate for their particular vulnerabilities. This is especially likely in the context of the NAO robot case study, as its design is intended to adjust its user's behaviour, albeit in ways that are intended to be a benefit for the child [53].

## 4   Ethical Perspectives on Social Robots

Through a theoretical lens, the ethics of HRI offer additional context to the operation of informed consent. This section discusses what is *right* and *wrong* in developing, deploying, and interacting with robots, outlining why it is important to delve into informed consent, and highlighting key problem areas that need to be considered. These discussions complement our prior exploration; by combining legal ideas of informed consent with ethical concerns and recommendations, we can obtain a more rounded and practical perspective for ensuring safe and ethical interaction between people and robots.

We present our exploration of existing ethical discussions surrounding social robots, organized into three broad categories. The first category considers how robots can influence and deceive people, mainly focusing on the robot's own nature and abilities. The second category focuses on the human perspective and the ethical implications of how humans respond to that influence. Finally, we examine works that attempt to resolve ethical unknowns regarding robots by understanding them in terms of metaphors with other entities. At each step of this exploration, we apply our lens of informed consent to critique these discussions, beginning a synthesis of legal and ethical perspectives that we expand on in Section 5.

## 4.1 Robotic Influence and Deception

Ethical scholarship in the HRI domain examines the different forms of influence that robots can wield over people, particularly when this influence is deceptive in nature. This angle of critique often employs descriptive frameworks to organize and classify different robot behaviours and impacts. This understanding provides a basis with which to frame robots' actions regarding potential concerns around informed consent.

To begin, consider works that look generally at the impact of technology on users. Tromp et al. [62] offers a framework for how technological artifacts may impact their users. This scholarship offers two axes to categorize objects' ability to influence users. These axes are 1) *force,* the strength of the influence, ranging from passive suggestions to mandatory; and 2) *salience*, the extent to which the decision is perceived by the user as internally versus externally motivated. What these axes demonstrate is a concern with 1) the extent to which an artifact deprives the user of their autonomy; and 2) the user's own perceived level of autonomy. This potential for dissonance between deprivation and perception of autonomy is central to the question of informed consent. This framework provides a useful connection between informed consent and HRI, since it is primarily concerned with how humans can be influenced, with or without their knowledge, by the objects they interact with.

For robots specifically, Shim & Arkin's [56] taxonomy of robot deception takes a descriptive approach to the issue of one form of influence: deception. This framework classifies robotic deception by target (human or non-human); beneficiary (robot or target); and means (whether deceiving through passive physical traits or active behaviour). By deconstructing deception into three binary dimensions, this framework provides a starting point for understanding the different ways that robots can manipulate the character and quality of information that a person holds, potentially compromising the validity of informed consent.

Danaher's [11] taxonomy of deception goes a step further, and attempts to articulate an ethical hierarchy of deception. This framework separates robot deception into external state deception (lies about the state of the world), and internal state deception (lies about the robot's capabilities). External state deception, in Danaher's view, is unexceptional, as it is ethically equivalent to human lying. Internal state deception is more complicated and can be further divided into superficial state deception (i.e., the robot suggests it has a capacity which it does not) and hidden state deception (i.e., the robot suggests it does not have a capacity that it actually does). In this framework, hidden state deception poses the largest ethical problem due to its potential for betrayal: it can manipulate through concealed abilities.

This approach conceptualizes robotic betrayal as a type of harmful deception unique to HRI. Through this lens, robot deception is a unique problem when it withholds information in a way that betrays a person's expectations of the robot (hidden state deception), which goes beyond the presentation of false information (superficial state deception). For example, when a robot makes a display of averting its painted eyes but continues observing a person through a hidden camera on the side of its head, this deception is considered a betrayal. However, if those superficial painted eyes are not accompanied by any actual visual capabilities to begin with, this incongruency is not considered a betrayal. The difference in these cases arises from whether or not a person is left vulnerable by their own expectations [11]. In the former case, a person's expectation that they are no longer being watched could lead to them letting their guard down, whereas a false belief that they *are* being watched does not take undue advantage of differing expectations. In concealing its real potential dangers, a robot denies users the information necessary to consent to interaction.

Danaher's conception of betrayal highlights a connection between the perceived ethics of a social robot and its relationship to human decision-making [11]. In this framework, the key point is that robot deception is unacceptable if it creates the potential for incongruency between humans' perceptions and their subsequent decision-making. This decision-making ability is of paramount importance in betrayal. The ability of a user to make decisions based on the maximum amount of information, even if it results in an over-estimation of the robot's capabilities, is the paramount consideration in whether a betrayal is occurring. Superficial state deception, then, is viewed as less concerning because it would, at most, result in overcautious decision-making, flowing from perceiving a robot as doing more than it really is.

This framework offers a parallel to informed consent. Both are concerned with humans' ability to make decisions with the information they are given. However, Danaher's framework fails to adequately conceptualize how robots' other deceptions may affect humans. For instance, external deception about facts may appear less concerning, but humans may infer that the robot is accessing the internet to gain information and is therefore less likely to provide false information than human memory. Superficial state deception can potentially produce an over-estimation of a robot's abilities, leading to over-trust and therefore impaired decision-making [8]. Danaher's [11] framework offers a first step towards employing informed consent to safeguard human autonomy in HRI.

### 4.2 Human Responses to Robot Sociality

Another major focus of ethical scholarship of social robots is on the impacts and responses that they effect in people. Where the previous section looked at works that were broadly descriptive in nature, these works are often more prescriptive: they identify outcomes in users that they deem to be ethically unacceptable.

Sparrow & Sparrow [61] present one such perspective. Their work largely concerns itself with care robots, specifically ones deployed in eldercare. Their reasoning proceeds in three parts: (1) social robots cannot provide genuine emotional affect; (2) humans have a moral duty to perceive the world accurately; and (3) humans desire to be cared about. In essence, the goal of a social robot is to simulate genuine affect (3), which it cannot do (1). Therefore, the only successful use of a social robot would be an act of deception (2). Through this lens, the feeling of being cared for by a social robot is an unacceptable delusion.

Sharkey & Sharkey [55] expand upon this idea, situating these issues in human dignity and the deceit of elders. They posit that the possibility of elders engaging with robotic babies or animals as though they were living beings would infringe their dignity. These arguments build upon a deep-rooted discomfort of incongruency between knowledge and reality, particularly when it leads to feelings of interpersonal closeness in human users.

This body of scholarship reiterates issues raised in Section 2 regarding parasociality and unilateral emotional responses. Further, it connects with informed consent, emphasizing the ways in which people are necessarily misinformed by their interactions with social robots. If human dignity and ethical judgments are prioritized, it seems unlikely that any standard of informed consent will satisfy the concerns that arise in an HRI context.

### 4.3 Understanding Through Metaphor

Faced with the considerable ethical unknowns surrounding social robots, many scholars have attempted to understand them through metaphors to other more familiar entities.

One common approach is to understand robots by analogy to humans. For example, the theory of ethical behaviorism [11] advocates for robots to be judged entirely by their external behaviours; similar arguments have been articulated as ethics of care [67] and environmental hypothesis [34]. These scholars argue that since inner states are ultimately unknowable in human-human interactions, it is unnecessary to insist upon emotional congruency in human-robot interactions. Further, these arguments, from a theoretical standpoint, reject the urge to anthropomorphize robots by assigning them thoughts and feelings.

Such anthropomorphization is at the heart of what Richards & Smart [46] call the android fallacy, which refers to the urge to anthropomorphize and infer mental states of robots in legal contexts, informed by the fantastical representations of robotics present in science fiction. They argue that conceiving of robots as having free will, as well as making assumptions about a robot's capabilities based on appearance and form, muddies legal waters for robots. Shim & Arkin's [57] binary categories of deception connotes some degree of android fallacy: they conceive of a robot as capable of lying for its own benefit, even though self-interest requires a sense of self which robots do not possess. Therefore, some might argue, the law should guard against commission of the android fallacy by legislating on the basis of what robots are, not based on what they seem to be. Kate Darling [12] pushes back against this recommendation, noting that since the public will anthropomorphize robots, it is in the public interest to legislate them with an attention to this perception [12, 46]. If being adequately informed about a robot's true nature is insufficient for people to treat that robot in accordance with that true nature, it is dangerous to assume people have real autonomy in any human-robot interactions.

The android fallacy is fundamentally concerned with how metaphor is being used to discuss robots. Metaphor framings are abundant in the social robotics scholarship. Sparrow & Sparrow's [61] and Sharkey & Sharkey's [55] inherent deception lens views robots as analogous to friends or loved ones, roles for which we typically expect congruency between their internal and external affects towards us. By contrast, external behaviour scholars articulate robots as being analogous to employees. Meacham & Studley [34] note that there are many professional caregiving jobs, like nursing, where we do not insist [46] on congruent emotional states; instead, we desire for behaviour to remain constant irrespective of internal emotional states. As Richards & Smart caution, however, the subtle shifts towards metaphor do a disservice towards accurately understanding the complexity of the issue.

Rather than attempting to choose between these frameworks and metaphors, each having its inevitable weaknesses, this section has attempted to extract the commonalities between these theories. They are not themselves holistic guides to the ethics of social robots, but rather lenses from which to draw inspiration and identify what is novel or unique. We observed that the fundamental questions circling the existing scholarship regard human autonomy, the internalizing of knowledge, and anxieties towards compromised decision-making. Such questions of autonomy, knowledge, and decision-making return our analysis to the informed consent frame.

### 4.4 Case Study Reflection: NAO Robot for ASD

We again leverage the NAO robot for ASD example in our examination of the ethical ramifications of social robots [53] as a child's interaction with the robot in the Shamsuddin et al. case study provides a fertile ground for robotic influence and deception. As proposed by Danaher [11], the hidden-state deception of the robot is likely the most probable to cause harms to the child. However, superficial state deception, with which a robot suggests it has a capacity that it does not [11], may be of particular note in the context of the robot's interactions with children. While it may be generally true that, as mentioned above, superficial state deception is less concerning due to its tendency to cause overcautious decision making, this type of deception may be more harmful for the child as robots may passively, through extended social interactions, impart upon children the idea that they are forming a social relationship that they are in no way capable of genuinely forming. This may well result in inaccurate perceptions of a relationship, leading to unproductive parasocial relationships between the robot and its user.

This concept of inappropriate parasocial relationships can be analogized to the position taken by Sharkey & Sharkey [55], who describe a loss of dignity in elders resulting from their interacting with robots as if they were human. Children are no less worthy of dignity, and no less likely to be at risk of just that loss of dignity, as noted by Hartzog [20], who demonstrated that children have been shown to consider robots to be their friends.

While prioritizing this framework of dignity in relation to human-robot interaction can lead to more positive outcomes, social robots like this one may also be viewed through the lens of metaphor. One potential point of view that may hold promise in mitigating deception and resulting harm is ethical behaviourism [11]. By judging the robot solely by its external behaviours (i.e., teaching robotically), anthropomorphization and Richards & Smart's android fallacy [46] may be avoided. Thus, it may be possible to, at the very least, reduce the likelihood of unhealthy relationships forming between the robot and its users by reshaping our perceptions of it with metaphor.

## 5 Achieving Informed Consent with Social Robots

Social robots present issues through technical, legal, and ethical lenses. Interactions with social robots are unlike interactions with other technological or living entities, be they asocial machines, virtual social agents, or people. This differentiation precludes us from treating robots the same as some other, more familiar category. The law provides rules governing the explicit and implicit obligations that present in interactions between people and through machines, but that these principles can be difficult to apply to the unique circumstances of social robots. Large bodies of scholarly work aim to understand and prescribe ethical interactions with robots.

While such work can inform the ethical development of HRI, it can be difficult to view it holistically when making practical applicational decisions. As this analysis demonstrates, these perspectives can conflict with one another, and occasionally with themselves. Specifically, we see a lack of literature attempting to address informed consent in HRI from a comprehensive, human-centered lens. To this end, we identify two areas of key concern that must be addressed when considering the validity of informed consent to an interaction with a social robot: congruence and human autonomy (Figure 1).

### 5.1 Congruence

Informed consent, as the name implies, requires that a user has the appropriate information necessary to make a decision, which is challenged when a robot conceals or implies false information about its nature. When describing an interaction with a social robot, we use the term *congruence* in two ways: to refer both to the alignment of a robot's outward appearance with its underlying capabilities (*design congruence*), and to the resulting alignment of the user's internal understanding of the interaction with its technical reality (*perceptual congruence*).

#### 5.1.1 Design Incongruence Leads to Perceptual Incongruence

As developed above, social robots necessarily deceive humans. Their basic concept relies on prompting users to behave toward them as they would a living, conscious being. They do this by feigning genuine social understanding and often an internal emotional state. Additionally, their physical and sensory capabilities are largely untethered to their outward appearance [27]. These incongruities in the robot's design lead directly to incongruities in the user's perception. Even when users are clearly informed about the robot's true nature in advance, the system can still utilize a person's biological instincts [59] to instill this false impression. This distortion is most strongly expressed in parasocial relationships, where the user not only imagines an immediate emotional response in the robot, but a long-term bond that is sustained across interactions. Though not always to this degree, some amount of design incongruence, and thus perceptual incongruence, is necessary in all social robots.

#### 5.1.2 Perceptual Incongruence and Informed Consent

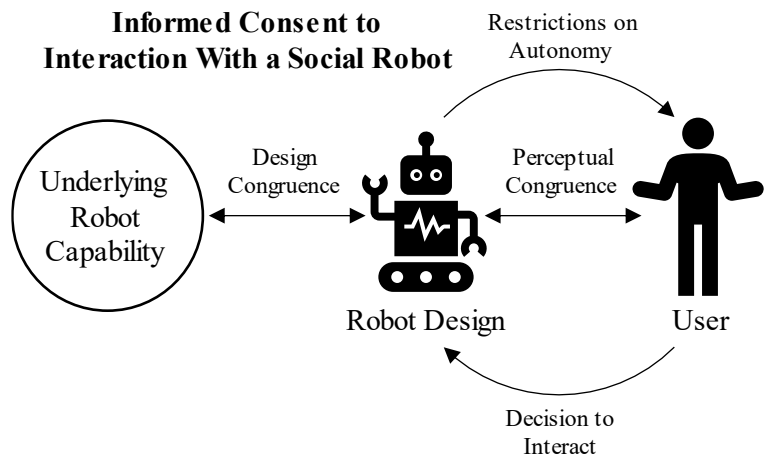

**Fig. 1:** A diagram illustrating the factors that contribute to a user's informed consent in their decision to interact with a social robot. A robot's design determines the congruence between a user's understanding of the robot and its true nature, while also introducing restrictions or manipulations on a user's ability to make an autonomous decision to interact.

Perceptual incongruence is an inherent feature of interaction with a social robot. This feature is at odds with ensuring that any user has full insight into the nature of the situation. Informed consent, as the term implies, relies on the notion that the user be properly informed about the situation before making a decision. Importantly, informed consent requires not only that information be provided to users, but that they understand and internalize it [44].

*5.1.3 Mitigating the Issue*

It may seem that social robots are simply incompatible with informed consent. Yet this observation offers little in terms of recommendations for designing more ethical systems. Instead, designers must attempt to balance the perceptual incongruence produced by their designs against the other benefits of the system.

One mitigation strategy may be to make the robot's deceptive nature as transparent to the user as possible. While this will not eliminate the incongruence, it at least provides as much information to the user as it can. It is also important to consider the scope of the incongruence: it may be possible to limit incongruence to areas that would have little relevance to the user's consent decision.

Such considerations must be made on a case-by-case basis, as different robots necessitate different degrees of perceptual incongruence in their users. By keeping this factor in mind during the design process, however, efforts can at least be made to mitigate the harm, protecting the validity of informed consent as much as possible.

**5.2 Human Autonomy**

Beyond misinforming users, social robots can further complicate informed consent by manipulating and restricting users' ability to make meaningful, independent decisions altogether. As with many technological designs, the overall design of a robot can coax and constrain users toward one choice or another, with or without them being aware. Given a robot's physical and social presence, the user's autonomy can be further diminished by leveraging their latent biases to encourage them to willingly cede their decisions to the robot out of both trust and convenience. The robot's dual role as social peer and calculating computer can create a uniquely authoritative impression that compromises the validity of informed consent.

*5.2.1 Automation Bias*

While much of our discussion has focused on whether robots pass as overly human, decision-making is one area where humans seem content to allow robots to live up to their full inhuman potential. The crux of the issue is that humans view algorithmic decision-making as less biased and therefore more reliable than the infinite nuance and bias of human thoughts [9, 40].

One of the common solutions to resist over-reliance on automated decision-making is to maintain a human in the loop. However, that is an oversimplified solution to a complex problem [9]. Often, relying on computer-generated outcomes is simply convenient; when human decisions vary so widely between people, why not default to a machine's outcome? One study found that people were heavily influenced by conformity bias, a desire to conform with initial opinions, when presented with both human and AI opinions [8]. The status of social robots as AI decision-makers affords them a level of authority distinct from human equivalents.

*5.2.2 Social Trust*

While automation bias is not unique to social robots, what is unique is social robots' ability to combine that bias with the trust and obligations afforded to a social peer. When a regular, asocial computer offers a suggestion, a user can freely ignore the decision without any feeling of direct social consequence. Advice and requests from social robots, however, with their physical presence and social status, demand a greater degree of consideration and response [2, 25]. This influence is further developed by the ability to take advantage of users' emotional vulnerabilities. By feigning emotional vulnerability [38] and inducing the development of parasocial relationships in the minds of users [39], a social robot can earn their intimate trust and encourage them to take actions they otherwise would not [24]. As a possibly extreme example, if a human were to incite another into committing a crime, then that person could be charged with counselling an offence [42]. However, in the case of an autonomous social robot, the waters become slightly muddied, with questions about imposing liability on the manufacturers, the users, and even the robot itself [23].

*5.2.3 Human Autonomy and Informed Consent*

The harm generated by losing aspects of human autonomy is twofold. First, reliance on automated decision-making on the basis that it is objective, neutral, and accurate obfuscates the nature of these systems, once again leaving users less informed. Algorithmic decision-making is based on data that may be biased, relying on stereotypes and correlation rather than nuanced evaluations of a given situation [9]. Second, eroding aspects of active human choice could lead to an erosion of human ability to make ethical decisions more broadly.

In Tromp et al.'s framework of influence, hidden influence occurs where users understand their behaviour as being internally motivated, rather than externally mediated or prompted [62]. This hidden influence could lead to what Ian Kerr termed the automation of virtue [28]. Kerr positions community as predicated on a shared group etiquette and morality, which is consistently reified by the members' decisions to adhere to this set of social norms. This community-building through choices is eroded, Kerr posits, when behaviour becomes automated through mediation, such as hidden influence [28]. Through this sort of influence, decisions we make about whether to break a social custom become pre-programmed behaviours selected for us, by our technology.

In restricting and funneling a user toward particular decisions, influencing their thoughts and behaviours both transparently and in ways they may not notice, it becomes difficult to describe a user's decision to consent as meaningful. No matter how informed the user is, it cannot be consent if the decision was not truly theirs to begin with.

*5.2.4 Preserving Human Autonomy*

As with the issue of congruence, responding to the issue of human autonomy must focus on awareness and restraint rather than decisive, absolute action. It is inevitable that technologies influence their users' behaviour, oftentimes in unexpected ways [62, 63, 66]. Rather than trying to avoid this outright, the solution must be to strive to anticipate these effects and identify those which are most damaging to informed consent. Most forms of influence may be entirely benign, but it is essential for designers to avoid developing systems that compromise their users' decisions about whether to interact with the system in the first place.

**5.3 Case Study Reflection: NAO Robot for ASD**

Our case study of a social robot for ASD children highlights especially perilous issues relating to informed consent. Incongruence between design and capabilities is common in social robots [48, 51] and can create a situation where truly informed consent is prohibitively difficult. While some mitigation strategies were suggested in section 5.1.3, the case of a robot for children with ASD may make those strategies ineffective. The ability to explain the complexities of the inner workings of a machine like this robot may mitigate some perceptual incongruency between the parents/guardians and the robot, but the child may well have a much more difficult time grasping these concepts, thus making this strategy less promising for socialization of children with ASD.

Further, issues relating to social trust may not be entirely clear in relation to this robot. Given the risk of forming parasocial relationships with social robots, these robots can induce individuals to take actions that they otherwise would not. This is further complicated by the fact that behavioral modification is part of the goal of a social robot for children with ASD. The extent to which this modification takes place must be carefully monitored, but this conflict highlights an important point: the nature of the issues and benefits surrounding social robots must be considered not with a broad, one-size-fits-all solution in mind, but with carefully tailored, contextual frameworks that address the specific needs of the users.

# 6 Conclusion

Ultimately, any path forward to integrating social robots into people's lives must be with a focus on adjusting to these new interaction paradigms in ways that protect people's rights. It is not as straightforward as simply explaining interaction to people, as we have demonstrated through this paper how informing people is fraught with challenges. Social robots offer a novel kind of interaction, strongly leveraging people's propensity toward anamorphism that is difficult, if not impossible, to ignore. Existing legal frameworks help to spotlight key challenges of knowledge of identity, meeting of the minds, and social obligations relating to social robots, and yet, law does not yet provide clear solutions given the novel ontological space of social robots. Ethical explorations of social robots have highlighted the need for a broader understanding of humans' propensity for manipulation and deception at the hands of anthropomorphized social robots capable of concealing their advanced capabilities, further complicating the issue of expecting people to readily understand a robot's intentions and capabilities.

As a step toward resolving the issue of informed consent when interacting with a social robot, we propose two important design goals. First, we should focus on mitigating incongruence between a robot's capabilities and what people perceive their capabilities to be. Second, we should focus on designing robots to maintain true human autonomy of interaction in lieu of potentially deceptive behavior. By focusing on these two goals, we envision that we can move closer toward a future where people truly are informed about the reality of the robots they interact with and can clearly consent to their interactions with these robots.

## Declarations

**Funding**

This research was funded under the Natural Sciences and Engineering Research Council of Canada (NSERC) Discovery Grants program (RGPIN-2024-05476) and the University of Manitoba Faculty of Science Interdisciplinary Research Grant. It was further supported by the NSERC Canada Graduate Scholarships – Master's scholarship, the University of Manitoba Graduate Fellowship, and the University of Manitoba Faculty of Science Undergraduate Student Research Award. This funding aside, the authors have no competing interests to declare that are relevant to the content of this article.

**Data Availability Statement**

We do not analyse or generate any datasets, because our work proceeds within a theoretical approach.

## References

[1] Henri François d'Aguesseau, Robert Pothier, Guillaume Le Trosne, and William Evans. 1853. *A treatise on the law of obligations, or contracts: translated from the French, with an introduction, appendix, and notes illustrative of the English law on the subject, by William David Evans* (3rd ed.). R.H. Small, Philadelphia.

[2] Wilma A. Bainbridge, Justin W. Hart, Elizabeth S. Kim, and Brian Scassellati. 2011. The Benefits of Interactions with Physically Present Robots over Video-Displayed Agents. *Int. J. Soc. Robot.* 3, 1 (January 2011), 41–52. https://doi.org/10.1007/s12369-010-0082-7

[3] Christoph Bartneck, Michel van der Hoek, Omar Mubin, and Abdullah Al Mahmud. 2007. "Daisy, daisy, give me your answer do!" switching off a robot. In *2007 2nd ACM/IEEE International Conference on Human-Robot Interaction (HRI)*, March 2007. 217–222.

[4] Allison S. Bohm, Edward J. George, Bennett Cyphers, and Shirley Lu. 2018. Privacy and Liberty in an Always-On, Always-Listening World. *Sci. Technol. Law Rev.* 19, 1 (January 2018). https://doi.org/10.7916/stlr.v19i1.4753

[5] Serena Booth, James Tompkin, Hanspeter Pfister, Jim Waldo, Krzysztof Gajos, and Radhika Nagpal. 2017. Piggybacking Robots: Human-Robot Overtrust in University Dormitory Security. In *Proceedings of the 2017*

*ACM/IEEE International Conference on Human-Robot Interaction* (*HRI '17*), March 06, 2017. Association for Computing Machinery, New York, NY, USA, 426–434. https://doi.org/10.1145/2909824.3020211

[6] Cynthia Breazeal. 2003. Toward sociable robots. *Robot. Auton. Syst.* 42, 3–4 (March 2003), 167–175. https://doi.org/10.1016/S0921-8890(02)00373-1

[7] William J. Brown. 2015. Examining Four Processes of Audience Involvement With Media Personae: Transportation, Parasocial Interaction, Identification, and Worship. *Commun. Theory* 25, 3 (2015), 259–283. https://doi.org/10.1111/comt.12053

[8] Federico Cabitza. 2019. Biases Affecting Human Decision Making in AI-Supported Second Opinion Settings. In *Modeling Decisions for Artificial Intelligence* (*Lecture Notes in Computer Science*), 2019. Springer International Publishing, Cham, 283–294. https://doi.org/10.1007/978-3-030-26773-5_25

[9] Carlo Casonato. 2021. AI and Constitutionalism: The Challenges Ahead. In *Reflections on Artificial Intelligence for Humanity*, Bertrand Braunschweig and Malik Ghallab (eds.). Springer International Publishing, Cham, 127–149. https://doi.org/10.1007/978-3-030-69128-8_9

[10] Emily S. Cross and Richard Ramsey. 2021. Mind Meets Machine: Towards a Cognitive Science of Human–Machine Interactions. *Trends Cogn. Sci.* 25, 3 (March 2021), 200–212. https://doi.org/10.1016/j.tics.2020.11.009

[11] John Danaher. 2020. Robot Betrayal: a guide to the ethics of robotic deception. *Ethics Inf. Technol.* 22, 2 (June 2020), 117–128. https://doi-org.uml.idm.oclc.org/10.1007/s10676-019-09520-3

[12] Kate Darling. 2016. Extending legal protection to social robots: The effects of anthropomorphism, empathy, and violent behavior towards robotic objects. In *Robot Law*. Edward Elgar Publishing, 213–232. https://doi.org/10.4337/9781783476732.00017

[13] Daniella DiPaola and Ryan Calo. 2024. Socio-Digital Vulnerability. https://doi.org/10.2139/ssrn.4686874

[14] Elizabeth Dula, Andres Rosero, and Elizabeth Phillips. 2023. Identifying Dark Patterns in Social Robot Behavior. In *2023 Systems and Information Engineering Design Symposium (SIEDS)*, April 2023. 7–12. https://doi.org/10.1109/SIEDS58326.2023.10137912

[15] Nicholas Epley, Adam Waytz, and John T. Cacioppo. 2007. On seeing human: A three-factor theory of anthropomorphism. *Psychol. Rev.* 114, (2007), 864–886. https://doi.org/10.1037/0033-295X.114.4.864

[16] Denise Y. Geiskkovitch, Derek Cormier, Stela H. Seo, and James E. Young. 2016. Please continue, we need more data: an exploration of obedience to robots. *J. Hum.-Robot Interact.* 5, 1 (March 2016), 82–99.

[17] Ella Glikson and Anita Williams Woolley. 2020. Human Trust in Artificial Intelligence: Review of Empirical Research. *Acad. Manag. Ann.* 14, 2 (July 2020), 627–660. https://doi.org/10.5465/annals.2018.0057

[18] Maartje M. A. de Graaf. 2016. An Ethical Evaluation of Human–Robot Relationships. *Int. J. Soc. Robot.* 8, 4 (August 2016), 589–598. https://doi.org/10.1007/s12369-016-0368-5

[19] John Harris and Ehud Sharlin. 2011. Exploring the affect of abstract motion in social human-robot interaction. In *2011 RO-MAN*, July 2011. 441–448. https://doi.org/10.1109/ROMAN.2011.6005254

[20] Woodrow Hartzog. 2014. Unfair and Deceptive Robots. *Md. Law Rev.* 74, (2015 2014), 785.

[21] Frank Hegel, Claudia Muhl, Britta Wrede, Martina Hielscher-Fastabend, and Gerhard Sagerer. 2009. Understanding Social Robots. In *2009 Second International Conferences on Advances in Computer-Human Interactions*, February 2009. IEEE, 169–174. https://doi.org/10.1109/ACHI.2009.51

[22] Fritz Heider and Marianne Simmel. 1944. An Experimental Study of Apparent Behavior. *Am. J. Psychol.* 57, 2 (1944), 243–259. https://doi.org/10.2307/1416950

[23] Ying Hu. 2018. Robot Criminals. *Univ. Mich. J. Law Reform* 52, (2019 2018), 487.

[24] Kumju Hwang and Qi Zhang. 2018. Influence of parasocial relationship between digital celebrities and their followers on followers' purchase and electronic word-of-mouth intentions, and persuasion knowledge. *Comput. Hum. Behav.* 87, (October 2018), 155–173. https://doi.org/10.1016/j.chb.2018.05.029

[25] Younbo Jung and Kwan Min Lee. 2004. Effects of Physical Embodiment on Social Presence of Social Robots. In *Proceedings of the 7th Annual International Workshop on Presence*, 2004. Valencia, Spain, 80–87.

[26] Peter H. Kahn, Aimee L. Reichert, Heather E. Gary, Takayuki Kanda, Hiroshi Ishiguro, Solace Shen, Jolina H. Ruckert, and Brian Gill. 2011. The new ontological category hypothesis in human-robot interaction. In

*Proceedings of the 6th international conference on Human-robot interaction - HRI '11*, 2011. ACM Press, Lausanne, Switzerland, 159. https://doi.org/10.1145/1957656.1957710
[27] Margot E. Kaminski, Matthew Rueben, William D. Smart, and Cindy M. Grimm. 2016. Averting Robot Eyes. *Md. Law Rev.* 76, (2017 2016), 983.
[28] Ian Kerr. 2010. Digital Locks and the Automation of Virtue. In *"Radical Extremism" to "Balanced Copyright": Canadian Copyright and the Digital Agenda*, Michael Geist (ed.). Irwin Law, Toronto, 247–303.
[29] Ian R. Kerr. 2003. Bots, Babes and the Californication of Commerce. *Univ. Ott. Law Technol. J.* 1, (2004 2003), 285–324.
[30] M Kottow. 2004. The battering of informed consent. *J. Med. Ethics* 30, 6 (December 2004), 565–569. https://doi.org/10.1136/jme.2003.002949
[31] Martin Kratz. 2020. When is a Contract of Adhesion Unconscionable? *Slaw*. Retrieved October 21, 2024 from https://www.slaw.ca/2020/08/05/when-is-a-contract-of-adhesion-unconscionable/
[32] William M. Landes and Richard A. Posner. 1981. An economic theory of intentional torts. *Int. Rev. Law Econ.* 1, 2 (December 1981), 127–154. https://doi.org/10.1016/0144-8188(81)90012-0
[33] Jamy Li. 2015. The benefit of being physically present: A survey of experimental works comparing copresent robots, telepresent robots and virtual agents. *Int. J. Hum.-Comput. Stud.* 77, (May 2015), 23–37. https://doi.org/10.1016/j.ijhcs.2015.01.001
[34] Darian Meacham and Matthew Studley. 2017. Could a Robot Care? It's All in the Movement. In *Robot Ethics 2.0*. Oxford University Press, New York, 97–112. https://doi.org/10.1093/oso/9780190652951.003.0007
[35] Andrew N. Meltzoff, Rechele Brooks, Aaron P. Shon, and Rajesh P. N. Rao. 2010. "Social" robots are psychological agents for infants: A test of gaze following. *Neural Netw.* 23, 8 (October 2010), 966–972. https://doi.org/10.1016/j.neunet.2010.09.005
[36] Jon F. Merz and Baruch Fischhoff. 1990. Informed consent does not mean rational consent: Cognitive limitations on decision-making. *J. Leg. Med.* 11, 3 (September 1990), 321–350. https://doi.org/10.1080/01947649009510831
[37] Ajung Moon, Maneezhay Hashmi, H. F. Machiel Van Der Loos, Elizabeth A. Croft, and Aude Billard. 2021. Design of Hesitation Gestures for Nonverbal Human-Robot Negotiation of Conflicts. *ACM Trans. Hum.-Robot Interact.* 10, 3 (July 2021), 24:1-24:25. https://doi.org/10.1145/3418302
[38] Youngme Moon. 2000. Intimate Exchanges: Using Computers to Elicit Self-Disclosure From Consumers. *J. Consum. Res.* 26, 4 (March 2000), 323–339. https://doi.org/10.1086/209566
[39] Nurhafihz Noor, Sally Rao Hill, and Indrit Troshani. 2021. Artificial Intelligence Service Agents: Role of Parasocial Relationship. *J. Comput. Inf. Syst.* (September 2021), 1–15. https://doi.org/10.1080/08874417.2021.1962213
[40] Cathy O'Neil. 2017. *Weapons of math destruction: how big data increases inequality and threatens democracy* (First paperback edition. ed.). Broadway Books, New York, NY.
[41] O. O'Neill. 2003. Some Limits of Informed Consent. *J. Med. Ethics* 29, 1 (2003), 4–7.
[42] Parliament of Canada. 2024. *Criminal Code, Section 22(1)*. Retrieved October 21, 2024 from https://laws-lois.justice.gc.ca/eng/acts/C-46/section-22.html
[43] Elizabeth Phillips, Xuan Zhao, Daniel Ullman, and Bertram F. Malle. 2018. What is Human-like?: Decomposing Robots' Human-like Appearance Using the Anthropomorphic roBOT (ABOT) Database. In *2018 13th ACM/IEEE International Conference on Human-Robot Interaction (HRI)*, March 2018. 105–113.
[44] Thaddeus Mason Pope. 2019. Informed Consent Requires Understanding: Complete Disclosure Is Not Enough. *Am. J. Bioeth.* 19, (2019), 27.
[45] Daryl Pullman. 2001. Subject Comprehension, Standards of Information Disclosure and Potential Liability in Research. *Health Law J.* (2001), 113–117.
[46] Neil M Richards and William D Smart. 2016. How should the law think about robots? In *Robot Law*, Ryan Calo, A Michael Froomkin and Ian Kerr (eds.). Edward Elgar Publishing, Cheltenham, UK, 3–22. https://doi.org/10.4337/9781783476732.00007

[47] Paul Robinette, Wenchen Li, Robert Allen, Ayanna M. Howard, and Alan R. Wagner. 2016. Overtrust of robots in emergency evacuation scenarios. In *2016 11th ACM/IEEE International Conference on Human-Robot Interaction (HRI)*, March 2016. 101–108. https://doi.org/10.1109/HRI.2016.7451740

[48] Julia Rosén, Jessica Lindblom, and Erik Billing. 2022. The Social Robot Expectation Gap Evaluation Framework. In *Human-Computer Interaction. Technological Innovation* (*Lecture Notes in Computer Science*), 2022. Springer International Publishing, Cham, 590–610. https://doi.org/10.1007/978-3-031-05409-9_43

[49] Maha Salem, Gabriella Lakatos, Farshid Amirabdollahian, and Kerstin Dautenhahn. 2015. Would You Trust a (Faulty) Robot? Effects of Error, Task Type and Personality on Human-Robot Cooperation and Trust. In *Proceedings of the Tenth Annual ACM/IEEE International Conference on Human-Robot Interaction* (*HRI '15*), March 02, 2015. Association for Computing Machinery, New York, NY, USA, 141–148. https://doi.org/10.1145/2696454.2696497

[50] Elaheh Sanoubari, Stela H. Seo, Diljot Garcha, James E. Young, and Veronica Loureiro-Rodriguez. 2019. Good Robot Design or Machiavellian? An In-the-Wild Robot Leveraging Minimal Knowledge of Passersby's Culture. In *2019 14th ACM/IEEE International Conference on Human-Robot Interaction (HRI)*, March 22, 2019. IEEE, 382–391. https://doi.org/10.1109/HRI.2019.8673326

[51] Lena T. Schramm, Derek Dufault, and James E. Young. 2020. Warning: This robot is not what it seems! exploring expectation discrepancy resulting from robot design. *ACMIEEE Int. Conf. Hum.-Robot Interact.* Figure 2 (2020), 439–441. https://doi.org/10.1145/3371382.3378280

[52] Stela H. Seo, Denise Geiskkovitch, Masayuki Nakane, Corey King, and James E. Young. 2015. Poor Thing! Would You Feel Sorry for a Simulated Robot? A comparison of empathy toward a physical and a simulated robot. In *2015 10th ACM/IEEE International Conference on Human-Robot Interaction (HRI)*, March 2015. 125–132.

[53] Syamimi Shamsuddin, Hanafiah Yussof, Luthffi Idzhar Ismail, Salina Mohamed, Fazah Akhtar Hanapiah, and Nur Ismarrubie Zahari. 2012. Initial Response in HRI- a Case Study on Evaluation of Child with Autism Spectrum Disorders Interacting with a Humanoid Robot NAO. *Procedia Eng.* 41, (January 2012), 1448–1455. https://doi.org/10.1016/j.proeng.2012.07.334

[54] Amanda Sharkey and Noel Sharkey. 2021. We need to talk about deception in social robotics! *Ethics Inf. Technol.* 23, 3 (September 2021), 309–316. https://doi.org/10.1007/s10676-020-09573-9

[55] Noel Sharkey and Amanda Sharkey. 2010. Living with robots: Ethical tradeoffs in eldercare. In *Close Engagements with Artificial Companions*. John Benjamins, 245–256. Retrieved July 14, 2022 from https://www.jbe-platform.com/content/books/9789027288400-nlp.8.29sha

[56] Jaeeun Shim and Ronald C Arkin. 2013. A Taxonomy of Robot Deception and Its Benefits in HRI. In *2013 IEEE International Conference on Systems, Man, and Cybernetics*, 2013. 2328–2335. https://doi.org/10.1109/SMC.2013.398

[57] Jaeeun Shim and Ronald C Arkin. 2013. A Taxonomy of Robot Deception and Its Benefits in HRI. In *2013 IEEE International Conference on Systems, Man, and Cybernetics*, 2013. 2328–2335. https://doi.org/10.1109/SMC.2013.398

[58] Francesca Simion, Elisa Di Giorgio, Irene Leo, and Lara Bardi. 2011. The processing of social stimuli in early infancy. In *Progress in Brain Research*. Elsevier, 173–193. https://doi.org/10.1016/B978-0-444-53884-0.00024-5

[59] Francesca Simion, Elisa Di Giorgio, Irene Leo, and Lara Bardi. 2011. The processing of social stimuli in early infancy. In *Progress in Brain Research*. Elsevier, 173–193. https://doi.org/10.1016/B978-0-444-53884-0.00024-5

[60] Daniel J. Solove. 2013. Introduction: Privacy Self-Management and the Consent Dilemma. *Harv. Law Rev.* 126, 7 (2013), 1880–1903.

[61] Robert Sparrow and Linda Sparrow. 2006. In the hands of machines? The future of aged care. *Minds Mach.* 16, 2 (October 2006), 141–161. https://doi.org/10.1007/s11023-006-9030-6

[62] Nynke Tromp, Paul Hekkert, and Peter-Paul Verbeek. 2011. Design for Socially Responsible Behavior: A Classification of Influence Based on Intended User Experience. *Des. Issues* 27, 3 (2011), 3–19.

[63] Peter-Paul Verbeek. 2006. Materializing Morality. *Sci. Technol. Hum. Values* 31, 3 (May 2006), 361–380. https://doi.org/10.1177/0162243905285847
[64] Katie Watson. 2013. Teaching the Tyranny of the Form: Informed Consent in Person and on Paper. *Narrat. Inq. Bioeth.* 3, 1 (2013), 31–34.
[65] Adam Waytz, John Cacioppo, and Nicholas Epley. 2010. Who Sees Human?: The Stability and Importance of Individual Differences in Anthropomorphism. *Perspect. Psychol. Sci.* 5, 3 (May 2010), 219–232. https://doi.org/10.1177/1745691610369336
[66] Langdon Winner. 1986. Do Artifacts have Politics? In *The whale and the reactor: a search for limits in an age of high technology*, Langdon Winner (ed.). University of Chicago Press, Chicago, 19–39.
[67] Gary Chan Kok Yew. 2021. Trust in and Ethical Design of Carebots: The Case for Ethics of Care. *Int. J. Soc. Robot.* 13, 4 (July 2021), 629–645. https://doi.org/10.1007/s12369-020-00653-w
[68] James Young. 2021. Danger! This robot may be trying to manipulate you. *Sci. Robot.* 6, 58 (September 2021). https://doi.org/10.1126/SCIROBOTICS.ABK3479/ASSET/78116656-C3F4-4AD3-A100-FEE563752A9A/ASSETS/IMAGES/LARGE/SCIROBOTICS.ABK3479-F1.JPG
[69] James E. Young, JaYoung Sung, Amy Voida, Ehud Sharlin, Takeo Igarashi, Henrik I. Christensen, and Rebecca E. Grinter. 2011. Evaluating Human-Robot Interaction: Focusing on the Holistic Interaction Experience. *Int. J. Soc. Robot.* 3, 1 (January 2011), 53–67. https://doi.org/10.1007/s12369-010-0081-8
[70] Chung-En Yu and Henrique F. Boyol Ngan. 2019. The power of head tilts: gender and cultural differences of perceived human vs human-like robot smile in service. *Tour. Rev.* 74, 3 (January 2019), 428–442. https://doi.org/10.1108/TR-07-2018-0097
[71] Abaid Ullah Zafar, Jiangnan Qiu, and Mohsin Shahzad. 2020. Do digital celebrities' relationships and social climate matter? Impulse buying in f-commerce. *Internet Res.* 30, 6 (2020), 1731–1762. https://doi-org.uml.idm.oclc.org/10.1108/INTR-04-2019-0142
[72] Jakub Złotowski, Diane Proudfoot, Kumar Yogeeswaran, and Christoph Bartneck. 2015. Anthropomorphism: Opportunities and Challenges in Human–Robot Interaction. *Int. J. Soc. Robot.* 7, 3 (June 2015), 347–360. https://doi.org/10.1007/s12369-014-0267-6
[73] Jakub Złotowski, Hidenobu Sumioka, Friederike Eyssel, Shuichi Nishio, Christoph Bartneck, and Hiroshi Ishiguro. 2018. Model of Dual Anthropomorphism: The Relationship Between the Media Equation Effect and Implicit Anthropomorphism. *Int. J. Soc. Robot.* 10, 5 (November 2018), 701–714. https://doi.org/10.1007/s12369-018-0476-5
[74] 1965. *Halushka v University of Saskatchewan et al*.
[75] 1974. *Menow v Jordan House Ltd*.
[76] 1980. *Reibl v Hughes*.
[77] 1988. *Crocker v. Sundance Northwest Resorts Ltd.*
[78] *Donoghue v. Stevenson, [1932] A.C. 562 (H.L.)*.
[79] *Scammell v. Ouston, [1941] A.C. 251. (H.L.)*.
[80] *Thornton v. Shoe Lane Parking Ltd. [1971] 1 All E.R. 686 (C.A.)*.
[81] *Cundy v Lindsay, (1878) 3 App Cas 458 HL*.
[82] *Lewis v. Averay [1971] 3 W.L.R. 603*.
[83] Torts, "Nature of tort law" at HTO-1 (2020 Reissue). *Halsbury's Laws of Canada (Online)*.
[84] *Kenny v. Lockwood, [1932] O.R. 141*.
[85] *Uber Technologies Inc. v. Heller, 2020 SCC 16*.